\documentclass{ISPIV2019_Template_Paper_Class}
\usepackage{amsmath}
\usepackage{xcolor}
\usepackage{graphicx}
\usepackage{caption} 
\def\citeapos#1{\citeauthor{#1}'s (\citeyear{#1})}
\newsavebox{\foobox}
\newcommand{\slantbox}[2][0]{\mbox{%
		\sbox{\foobox}{#2}%
		\hskip\wd\foobox
		\pdfsave
		\pdfsetmatrix{1 0 #1 1}%
		\llap{\usebox{\foobox}}%
		\pdfrestore
}}

\newcommand\slant[2][.25]{\slantbox[#1]{$#2$}}
\let\oldbibliography\thebibliography
\renewcommand{\thebibliography}[1]{
  \oldbibliography{#1}%
  \setlength{\itemsep}{7pt}%
}
\usepackage{bm}
\DeclareSymbolFont{newfont}{OML}{cmm}{m}{it}
\DeclareMathSymbol{\Epsilon}{3}{newfont}{15}

\begin{document}
\title{Three-dimensional tracking of finite-size spheres in a turbulent boundary layer}
\author{Yi Hui Tee$^{1*}$, Diogo Barros$^2$, Ellen K. Longmire$^1$}
\affiliation{
		$^1$ University of Minnesota, Aerospace Engineering and Mechanics, Minneapolis, USA\\
 		\vspace{7pt}
	     	$^2$ Aix-Marseille {Universit\'e}, CNRS, IUSTI, Marseille, France\\
		\vspace{7pt}
		$^*$ teexx010@umn.edu
		}
\maketitle

\begin{abstract}
The motion of individual magnetic wax spheres with specific gravities of 1.003, 1.050 and 1.150 was investigated in turbulent boundary layers with $Re_{\scriptsize{\slant\tau}}=700$ and 1300 ($d^+ = 60$ and 120).
The spheres were marked with dots all over the surface to monitor their translation and rotation via high-speed stereoscopic imaging.
Upon release from rest on a smooth wall, each sphere typically accelerated strongly over a  streamwise distance of one boundary layer thickness before approaching an approximate terminal velocity.
Spheres with sufficient net upward force lifted off of the wall once released before descending back towards the wall.
These spheres mostly translated with the fluid above the wall, undergoing saltation or resuspension, with minimal rotation about all axes.
By contrast, spheres that did not lift off after release mainly slid along the wall. 
As they propagated downstream, they began to roll forward with occasional lift-off events of smaller magnitude.
All of the lift-off activities observed were limited to the buffer and logarithmic layers. 
Both translation and rotation of the spheres were significantly affected by the wall turbulence.

\end{abstract}
\section{Introduction}
Particle-laden turbulent flows occur in many applications ranging from industrial processes to the environment.
In a wall-bounded flow, the particle motion is complicated by interactions with both the turbulent fluid and the wall itself.
When a particle is larger than the smallest fluid eddies, it can experience variations in shear and normal forces around its circumference.
Additionally, wall friction and restitution effects can affect both the translation and rotation of particles. 
Depending on the surrounding conditions, particles can either collide with or lift off from the wall or slide or roll along it.
All of these effects can significantly impact particle suspension, deposition and transport.

Various early experiments examined particle dynamics in turbulent open-channel flows.
Among others, \cite{sutherland1967proposed} investigated how grains in a sediment bed were brought into motion by the fluid.
He proposed an entrainment hypothesis whereby strong turbulent eddies could disrupt the viscous sublayer and lift the grain off of the bed.
\cite{francis1973experiments} observed rolling, saltation and suspension behavior of heavy grains transported over a planar rough bed.
Meanwhile, \cite{sumer1978particle} observed that a sand-coated wax sphere with diameter of approximately 30 viscous units ($d^+$) propagated upwards and downwards repetitively throughout its trajectory. 

To better understand particle-turbulence interactions in wall-bounded flows, direct visualization techniques were incorporated in the research by \cite{kaftori1995particle}, \cite{ninto1996experiments} and \cite{van2013spatially} to name a few.
These studies concluded that particle resuspension and deposition events in the near wall region were strongly influenced by coherent flow structures. 
Specifically, \cite{van2013spatially} reported that in their time-resolved PIV and PTV experiments, all lift-off events of polystyrene beads ($d^+=10$) at friction Reynolds number, $Re_{\scriptsize{\slant\tau}}=435$ were due to ejection events generated by passing vortex cores and positive shear. 
Once lifted beyond the viscous sublayer, the particles either stayed suspended in the fluid or saltated along the wall depending on the type of coherent structures that they encountered. 

Simulations of finite-size particles in turbulent boundary layers are very limited.
\cite{picano2015turbulent} and \cite{costa2018effects}, among others, performed direct numerical simulations on dense suspensions of finite-size spheres in turbulent channel flow.
\textcolor{black}{Meanwhile, \cite{zeng2008interactions} focused on the fluid forces acting on a fixed sphere located above the wall with finite gap.
The mean lift forces simulated for spheres with $1.78\leq d^+/2\leq1 2.47$ centered at a wall-normal location $y^+=17.31$ from the wall at $Re_{\scriptsize{\slant\tau}}=178.12$ were negative in all cases.
In this context, tomographic PIV performed by \cite{van2018experimental} also suggested a negative lift contribution on a tethered sphere with $d^+/2=25$ centered at $y^+=43$ above the wall at $Re_{\scriptsize{\slant\tau}}=352$ due to the sphere wake tilting away from the wall.
Under both studies, the gaps between the wall and the bottom sphere were $4.84\leq \bigtriangleup y^+\leq 18$.}
On the other hand, \cite{hall_1988}, who measured the mean lift force acting on a stationary particle lying on the wall in a turbulent boundary layer, reported a positive lift contribution.
The experimental data showed that for $3.6 < d^+ < 140$ and particle Reynolds number, $6.5<Re_p<1250$, the normalized mean lift force was strongly positive and could be approximated by $F^+_L=(20.90\pm1.57)(d^+/2 )^{2.31\pm0.02}$.

Tracking the dynamics of discrete particles with significant size, though challenging, is fundamental to characterize the particle-fluid interactions.
To track the translation and rotation of spheres over a three-dimensional (3D) domain, current experimental methods include printing specific patterns over the sphere surface \citep[]{zimmermann2011tracking} as well as embedding visible tracers into the interior of transparent spheres \citep[]{klein2013simultaneous}.
The first method relies on comparing the unique pattern captured by high-speed cameras with the respective synthetic projections to extract the absolute orientation of the sphere; the latter method focuses on Lagrangian tracking of the injected tracers using three or more cameras and computing the rotation rate based on an optimized rotation matrix demonstrated by \cite{kabsch1976solution}. 
A recent method proposed by \cite{barros2018measurement} extended \citeapos{klein2013simultaneous}  methodology to opaque spheres.
This method is justifiable when particle-fluid index matching is not possible.
Small dots were marked all over the solid surface, and two cameras were employed for the 3D reconstruction.

In the present study, we conduct particle tracking experiments to investigate the 3D motion of individual finite-size spheres within a turbulent boundary layer. 
Multiple sphere densities and flow conditions are considered. 
To obtain both the translation and rotation of a sphere, \citeapos{barros2018measurement} methodology is adapted to the requirements of the current experimental setup. 
Based on the reconstructed sphere orientations and trajectories, both the translational and rotational kinematics of the spheres are investigated.

\vspace {-1.5mm}

\section{Methodology}
\vspace {-0.5mm}

The experiments were conducted in a recirculating water channel facility. 
The channel test section, which is constructed of glass, is 8 m long and 1.12 m wide. 
A 3 mm cylindrical trip-wire was located at the entrance of the test section to trigger the development of a turbulent boundary layer along the bottom wall. 
Hereafter $x$, $y$ and $z$ define the streamwise, wall-normal, and spanwise directions, respectively. 

\begin{figure}[]

        \centering
    	 \includegraphics[trim={0 2.45cm 0 1.1cm },clip,width=0.92\textwidth]{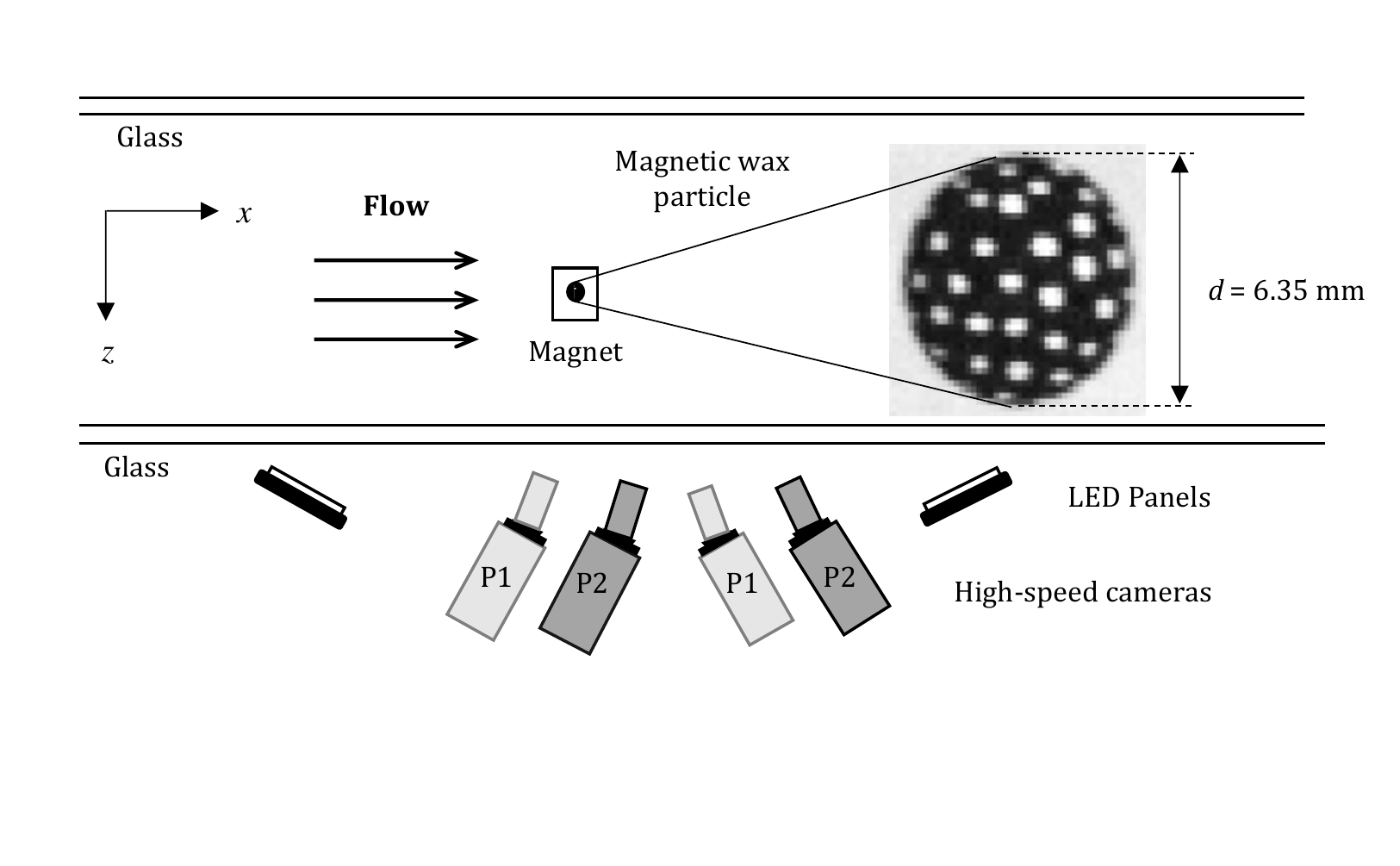} %
	 \caption{Top view of the experimental setup: two pairs of high-speed cameras (P1 and P2) were aligned in stereoscopic configuration for capturing the trajectory and rotation of a marked sphere over a long field of view. Inset: example of sphere captured in grayscale with diameter ($d$) spanning 43 pixels or 6.35 mm.}
		\label{fig:1}

\end{figure}

To achieve a repeatable and controllable initial condition, magnetic spheres molded from a mixture of wax and iron oxide were used. 
By controlling the amount of iron oxide, three spheres were fabricated with varying densities.
The black sphere surfaces were painted with white dots at arbitrary locations using an oil-based pen to monitor both translation and rotation (see figure \ref{fig:1}). 
For each run, a given sphere was held statically on the smooth wall in the boundary layer by a magnet placed flush with the outer wall of the channel. 
The sphere was positioned at a location 4.2 m downstream of the trip wire and 4 boundary layer thicknesses ($\slant\delta$) away from the nearest sidewall.
This location will be considered as the origin in $x$ and $z$, with the bottom wall as $y=0$. 
The magnet was then deactivated, releasing the sphere and allowing it to propagate with the incoming flow. 
A screen was located at the end of the test section to capture the sphere and prevent it from recirculating around the channel.

Two pairs of Phantom v210 high speed-cameras from Vision Research Inc. were arranged in stereoscopic configurations to track the sphere in 3D space over a relatively long field of view. 
The angle between the two stereoscopic cameras was set to approximately $30^\circ$ for both camera pairs.
All cameras were fitted with 105 mm Nikon Micro-Nikkor lenses with aperture $f/16$. 
Scheimpflug mounts were added to all cameras so that the images were uniformly focused across the fields of view. 
Prior to running the experiments, the optical system was calibrated by displacing a two-level plate (LaVision Type 22) across nine planes in the spanwise ($z$) direction for volumetric reconstruction. 
A third order polynomial fit was obtained for each plane from both cameras using the calibration routine of Davis 8.4 to generate the mapping function of the volumetric calibration. 
The root-mean-square error of the grid point positions was between 0.05 and 0.1 pixels indicating an optimal fit.
Image sequences were captured at a sampling frequency of 480 Hz with image resolution of 1280 x 800 pixels. Three white LED panels illuminated the domain considered.

To understand the effect of turbulence, the experiments were conducted at $Re_{\scriptsize{\slant\tau}}$ of 700 and 1300, respectively. 
This corresponded to free-stream velocities ($U_\infty$) of 0.22 m/s and 0.49 m/s, with $\slant\delta$ of 0.073 m and 0.066 m, respectively.
The water depth was maintained at 0.394 m under both fluid conditions. 
Here, the mean flow statistics of the unperturbed turbulent boundary layers were determined from planar PIV measurements in streamwise wall-normal planes.

Spheres with diameter ($d$) of 6.35 mm and specific gravities ($\slant\rho_s/\slant\rho_f$) of 1.003 (P1), 1.050 (P2) and 1.150 (P3) were considered, where $\slant\rho_s$ is the sphere density and $\slant\rho_f$ is the fluid density. 
The spheres were significantly larger than the Kolmogorov length scale, with $d^+$ of 60 and 120 respectively. 
The initial particle Reynolds numbers defined as $Re_p=U_{rel}d/\slant\nu$ were 800 and 1900, where $U_{rel}$ is the mean particle-fluid relative velocity at the particle center upon release, and $\slant\nu$ is the kinematic viscosity of the water. 
Details of the experimental parameters are summarized in Table \ref{tab:1}.
The particles were tracked over a streamwise distance up to $x = 6\slant\delta$.
For each case considered, $N=10$ trajectories were captured using the same sphere.

\begin{table}[]
\caption{Summary of experimental parameters. $|V_s|$ represents sphere settling velocity magnitude in quiescent flow; $\overline{F_L}$ and $F_b$ denote the mean lift force based on \citeapos{hall_1988} expression and the net buoyancy force, respectively.}
\centering
\small
\renewcommand*{\arraystretch}{1.11}
\setlength{\tabcolsep}{10pt}   
\begin{tabular}{c  c  c  c  c  c  c }
\hline\hline \\[-2.65ex]
$Re_{\scriptsize{\slant\tau}}$ & Initial $Re_p$ & $d^+$ & Sphere & $\slant\rho_s/\slant\rho_f$ & $|V_s|/U_\infty$ & Initial $\overline{F_L}/|F_b|$ \\[0.35ex]
\hline \\[-2.7ex]
700       & 800            & 60    & P1     & 1.003          & 0.07          & $14\pm2$     \\
          &                &       & P2     & 1.050          & 0.40          & $0.86\pm0.13$      \\
          &                &       & P3     & 1.150          & 0.74          & $0.28\pm0.04$      \\[0.35ex]
\hline  \\[-2.7ex]

1300      & 1900           & 120   & P1     & 1.003          & 0.03          & $70\pm10$     \\
          &                &       & P2     & 1.050          & 0.18          & $4.3\pm0.6$      \\
          &                &       & P3     & 1.150          & 0.33         & $1.4\pm0.2$     \\[0.35ex]
\hline\hline
\end{tabular}
\label{tab:1}
\end{table}

Before computing the particle translation and rotation, the grayscale images were first pre-processed using Matlab to isolate the sphere from the background. 
A standard circular Hough Transform routine was applied to locate the sphere.
Next, the background surrounding the sphere was removed by setting the intensity values to 0 (black).
The extracted sphere images were then imported to Davis 8.4.
Here, the images were further processed with 3 x 3 Gaussian smoothing and sharpening to increase the dot contrasts. 
Pixel intensity values that were less than the white dots were set to 0 to isolate the dots from the sphere image. 
Subsequently, a 3D-PTV routine was implemented to reconstruct the dot coordinates from both camera pairs based on the volumetric calibration mapping function.
The mean disparity error, $\Epsilon_{disp}^*$ calculated by projecting the 3D reconstructed markers back to the camera image was 0.8 px. 
This gives an estimate of the uncertainty in the marker locations due to reconstruction errors \citep{wieneke2008volume}.

The data sets obtained from PTV were composed of the 3D coordinates of true and ghost markers and their corresponding 3D velocity vectors. 
Hence, the filtering methodology proposed by \cite{barros2018measurement} was employed to remove the ghost tracks.  
Once the true markers had been determined, the sphere centroid was determined by applying the equation of a sphere. 
Then, a rotation matrix that best aligned the markers of consecutive images was obtained \citep{barros2018measurement}.
In all runs, different processing frequencies were used in tracking the markers. 
This was to ensure that between subsequent images, the markers would displace larger than the pixel uncertainty while staying within the camera field of view for at least one time step to avoid wrong marker pairing.
Here, the uncertainty of the translation and rotation displacement computed based on the r.m.s. between the raw data and the data smoothed by a quintic spline were 0.75 and 0.53 px respectively \citep{epps2010evaluating, schneiders2017track}.

\vspace {-1.5mm}

\section{Results and Discussion}
\vspace {-0.5mm}

\begin{figure}[]

        \centering
        \begin{subfigure}{\label{fig:2a}
        		\includegraphics[trim={2mm 2.3mm 2mm 2mm},clip,width=.486\textwidth]{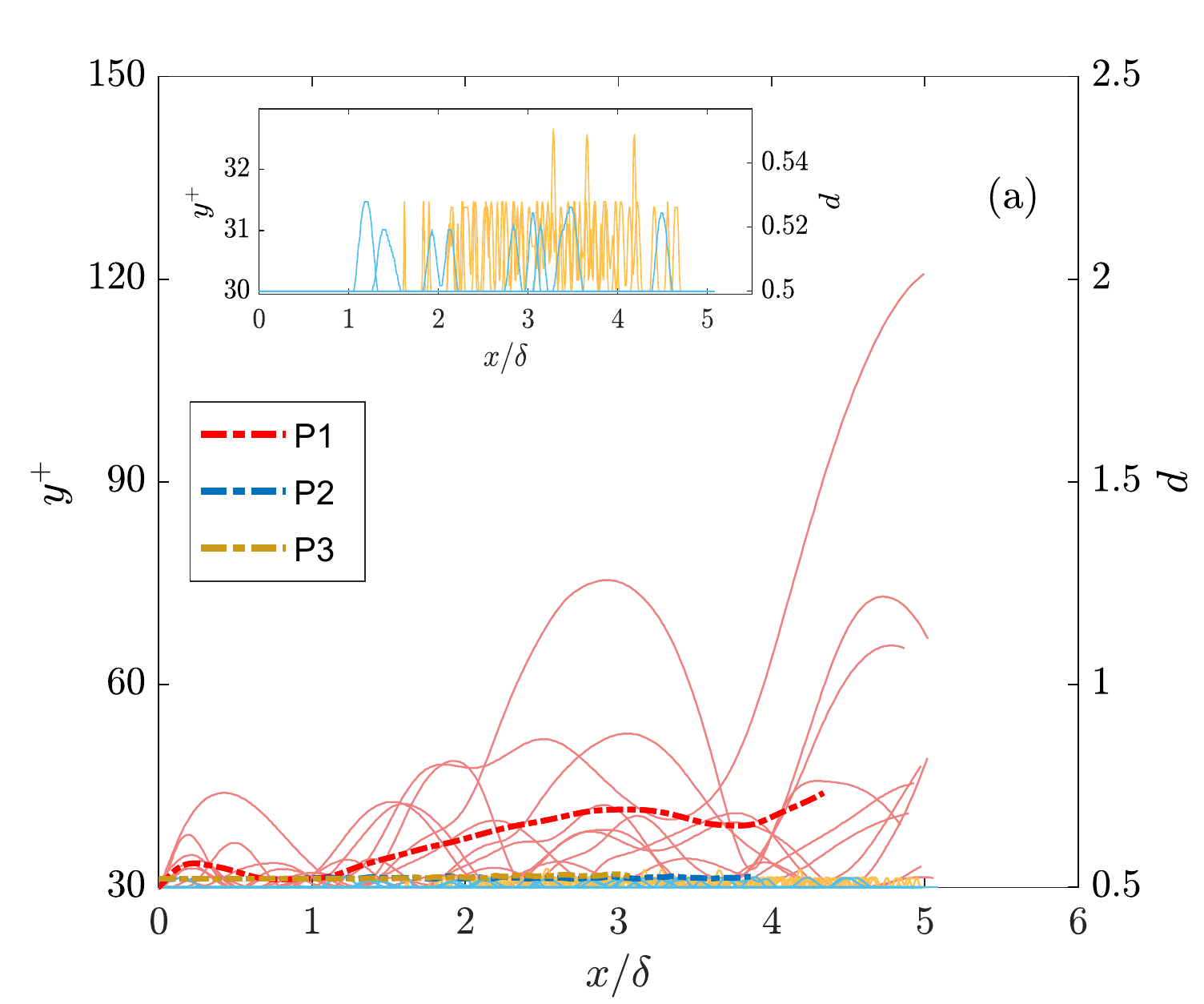}
	}
	 		\end{subfigure}	
	\begin{subfigure}{\label{fig:2b}
		\includegraphics[trim={0 1.5mm 0 0},clip,width=.478\textwidth]{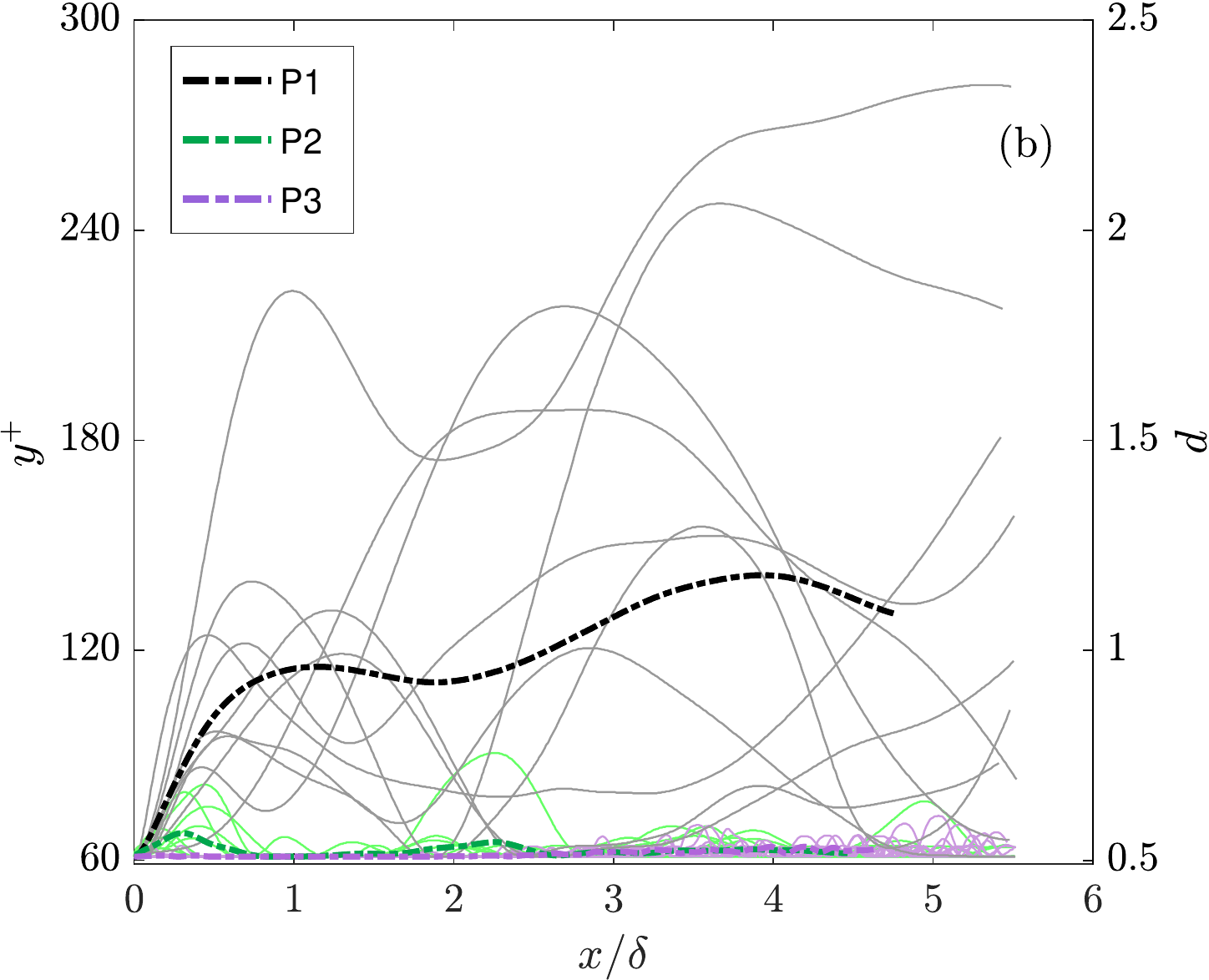}
	}
	\end{subfigure}
	\vspace {-3.5mm}
	\caption{Sphere wall-normal trajectories, $y$ at (a) $Re_{\scriptsize{\slant\tau}}=700$ and (b) $Re_{\scriptsize{\slant\tau}}=1300$ respectively plotted based on the centroid positions. Bold dashed lines represent the ensemble average of 10 runs, each of which is plotted as a solid line. Specific gravity: P1 = 1.003, P2 = 1.050 and P3 = 1.15. The inset in (a) illustrates a sample trajectory for P2 and P3.}
		\label{fig:2}
	\vspace{-0.2cm}

\end{figure}

Figure \ref{fig:2} shows the sphere wall-normal ($y$) trajectories plotted against their streamwise distance traveled. 
For this and subsequent figures, all particle runs are superposed within the same figure as solid lines. 
Bold dashed lines depict the ensemble average of position at a given time for each case where $\overline{\mathbf{X}(t_i)}=(1/N)\sum\limits_{j=1}^{N} \mathbf{X_j}(t_i)$, unless otherwise specified. 
The average plots are cut off at the time when $N$ falls below 10.
Firstly, we would like to focus on the initial sphere motion right after $x=0$. 
To complement our observations, the sphere mean lift forces were computed using \citeapos{hall_1988} equation obtained for a fixed sphere in a turbulent boundary layer where $\overline{F_L}=F_L^+\slant\nu^2\slant\rho_f$.
The uncertainty of this equation was approximately $\pm0.15\overline{F_L}$ for both fluid conditions investigated.
The sphere net buoyancy forces, $F_b = (m_s - m_f)g$, were also calculated (see Table \ref{tab:1}). 
Here, $m_s$ refers to the mass of sphere, $m_f$ defines the mass of fluid displaced and $g$ is the gravitational acceleration.
As all spheres are denser than water, the net buoyancy force will always act downward in the direction of gravity.
Thus, if  ${F_L} > |F_b|$, the sphere shall lift off upon release, and vice versa.

For the least dense sphere P1, at $Re_{\scriptsize{\slant\tau}} = 700$, $\overline{F_L} \cong 14F_b$. 
Increasing $Re_{\scriptsize{\slant\tau}}$ to 1300 doubled $d^+$ and thus the sphere mean lift force increased fivefold.
These estimations agreed very well with our observations where sphere P1 always lifted off of the wall once released at both $Re_{\scriptsize{\slant\tau}}$ (plotted as red in figure \ref{fig:2a} and as black in figure \ref{fig:2b}). 
Additionally, owing to the stronger resultant upward force, the average initial lift-off height of sphere P1 at $Re_{\scriptsize{\slant\tau}}=1300$ was three times higher than at $Re_{\scriptsize{\slant\tau}}=700$.
This shows that the initial lift-off height correlates strongly with the local mean shear.

For sphere P2, although $\overline{F_L} \cong 0.86F_b$ at $Re_{\scriptsize{\slant\tau}}=700$, the spheres lifted off of the wall twice (plotted as blue in figure \ref{fig:2a}). 
In the remaining runs, the spheres mainly translated along the wall once released.
Meanwhile, at higher $Re_{\scriptsize{\slant\tau}}$ (plotted as green in figure \ref{fig:2b}), the spheres lifted off of the wall eight out of ten runs due to the stronger upward force ($\overline{F_L} \cong 4.3F_b$).
In this context, the mean initial lift-off height of sphere P2 was only $14\%$ of that from P1 under the same flow condition.
For the densest sphere P3, at lower $Re_{\scriptsize{\slant\tau}}$, no initial lift-off was observed. 
The sphere did not have sufficient lift force to overcome the downward force and translated along the wall upon release.
At higher $Re_{\scriptsize{\slant\tau}}$, sphere P3 lifted off of the wall only once although $\overline{F_L} /|F_b| \cong 1.4$. 
These observations suggest that even though P2 and P3 are $5\%$ and $15\%$ denser than P1, the effect of net buoyancy on the initial sphere lift-off event is strong.

In general, most of our results on sphere initial lift-off events agreed well with \cite{hall_1988}.  
For a sphere lying on wall, if the lift force (positive) exceeds the net buoyancy force (negative), it lifts off from rest.
However, some disagreements were observed for sphere P2 at $Re_{\scriptsize{\slant\tau}}=700$ and sphere P3 at $Re_{\scriptsize{\slant\tau}}=1300$. 
This could be due to when considering the mean lift forces, fluctuating components were averaged out from the instantaneous forces.  
\textcolor{black}{In this context, the fluctuating fluid velocities ($u_{rms}/\overline{U_f}$) at the sphere initial centroid positions at $Re_{\scriptsize{\slant\tau}}=700$ and $1300$ were $0.2$ and $0.15$ respectively.
If we compute the instantaneous lift forces where $F_L/\overline{F_L}=(\overline{U_{f}}\pm u_{rms})^2 /\overline{U_{f}}^2$, the values vary up to $40\%$ and $26\%$ from $\overline{F_L}$ respectively. 
Hence, depending on the type of the coherent structures that the sphere has encountered locally, the mean lift force could either underestimate or overestimate the instantaneous lift force magnitude. 
This large variation will also explain the widespread in initial lift-off heights observed even though mean lift force and net buoyancy force remain unchanged for a same sphere at same $Re_{\scriptsize{\slant\tau}}$.}
It is also important to point out that \citeapos{hall_1988} mean lift force equation was fitted based on experimental data up to $Re_p=1250$. 
Hence, our observations on sphere P3 at $Re_{\scriptsize{\slant\tau}}=1300$ with $Re_p=1900$ suggest that \citeapos{hall_1988} equation might over-predict the mean lift force for spheres with $Re_p>1250$.

\textcolor{black}{Spheres that initially lifted off always descended towards the wall due to gravity after reaching a local maximum in height. 
For sphere P1 at $Re_{\scriptsize{\slant\tau}}=1300$, it either contacted the wall and then lifted off or else ascended to a higher location without returning to the wall.
In all other lifting cases, the spheres always contacted the wall and then either lifted off again or else translated along the wall.}

Regardless of their initial motions, in most runs, the spheres experienced multiple lift-off events including saltation or resuspension with lift-off angles less than $12^{\circ}$ throughout their trajectories. 
As sphere P1 has density closest to neutral buoyancy, it always lifted off to a higher location than other spheres, with maximum magnitude up to $2d$ above the wall. 
Also, this sphere could ascend to a greater height than its initial peak location as it propagated downstream.  
On the other hand, the lift-off magnitudes for spheres P2 and P3 stayed within half a diameter or less above the wall.
Interestingly, sphere P3 exhibited a consistent lift-off pattern that was different from the other spheres.
Although this sphere did not generally lift off upon release, as it traveled along the wall with the incoming fluid, repeated small lift-off events began to take place downstream. 
This behavior was frequently observed under both fluid conditions and occurred more consistently starting at $x\approx2\slant\delta$ when $Re_{\scriptsize{\slant\tau}} = 700$.

As reported by \cite{van2013spatially}, these lift-off events could be triggered by the ejection-sweep cycles.
In this context, the local pressure fields and velocity fields around the finite-size sphere are continually altered by wall turbulence as well as vortex shedding.
Therefore, the wall-normal trajectory of a sphere which is strongly governed by the instantaneous lift force will be affected by the type of coherent structures and vortices the sphere encounters at different locations.

\begin{figure}[]

        \centering
            \begin{subfigure}{\label{fig:3a}
    			 \includegraphics[trim={0 0.5mm 0 0.1mm},clip,width=.475\textwidth]{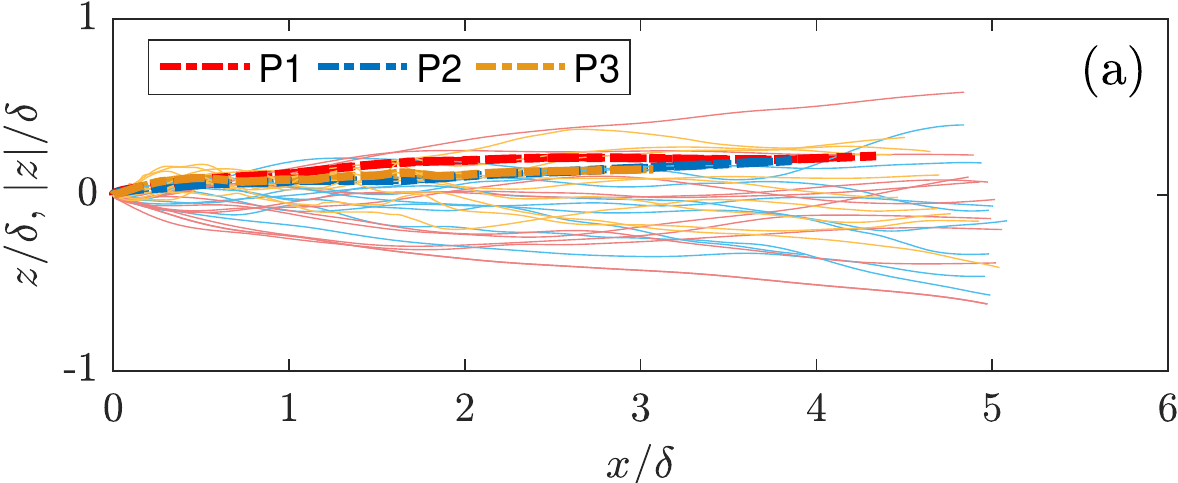}
    			 }
	 		\end{subfigure}	\hspace{0.5mm}	
            \begin{subfigure}{\label{fig:3b}
    			 \includegraphics[trim={0 0.5mm 0 0.1mm},clip,width=.475\textwidth]{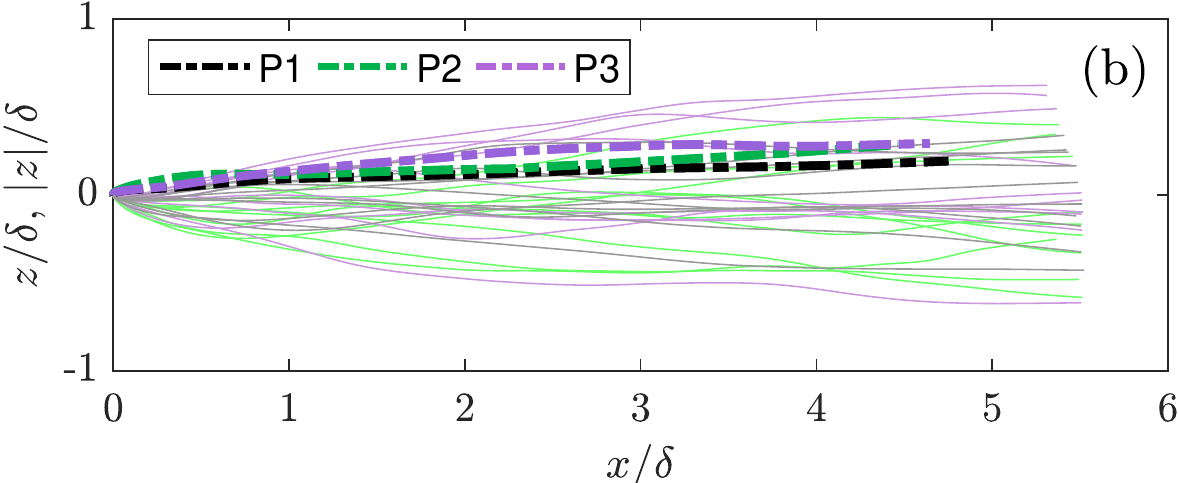}
	 		}
	 		\end{subfigure}
	 		    	 \vspace {-4mm}

	 		\caption{Solid lines - sphere spanwise trajectories, $z$ at (a) $Re_{\scriptsize{\slant\tau}}=700$ and (b) $Re_{\scriptsize{\slant\tau}}=1300$ respectively. Bold dashed lines - ensemble average of the absolute spanwise positions, $\overline{|z|}$.}
        \label{fig:3}
\end{figure}

The spanwise trajectories of the spheres are plotted in figure \ref{fig:3}.
Once released, instead of propagating along $z=0$, the spheres typically moved sideways.
Then, they either continued to propagate in one direction or else reversed and traveled in the opposite direction, with some crossings over $z=0$.
\textcolor{black}{For P3 at $Re_{\scriptsize{\slant\tau}}=700$, most of the initial spanwise curves fluctuated at higher frequency than other cases when $x<2\slant\delta$.}
This fluctuation depicts that the sphere changed direction repeatedly.
In all cases, the mean absolute spanwise migration angles (plotted as bold lines in figure \ref{fig:3}) lay between $3^\circ$ and $5^\circ$ from $z=0$. 
Meanwhile, the maximum spanwise migration distances were approximately 5$-7d$ or 8$-12\%$ of the streamwise distance traveled.
In most runs, the spanwise migration magnitudes were larger than the maximum lift-off magnitude.

\begin{figure}[]
        \centering
        \begin{subfigure}{\label{fig:4a}
        		\includegraphics[trim={0 1.6mm 0 0},clip,width=.48\textwidth]{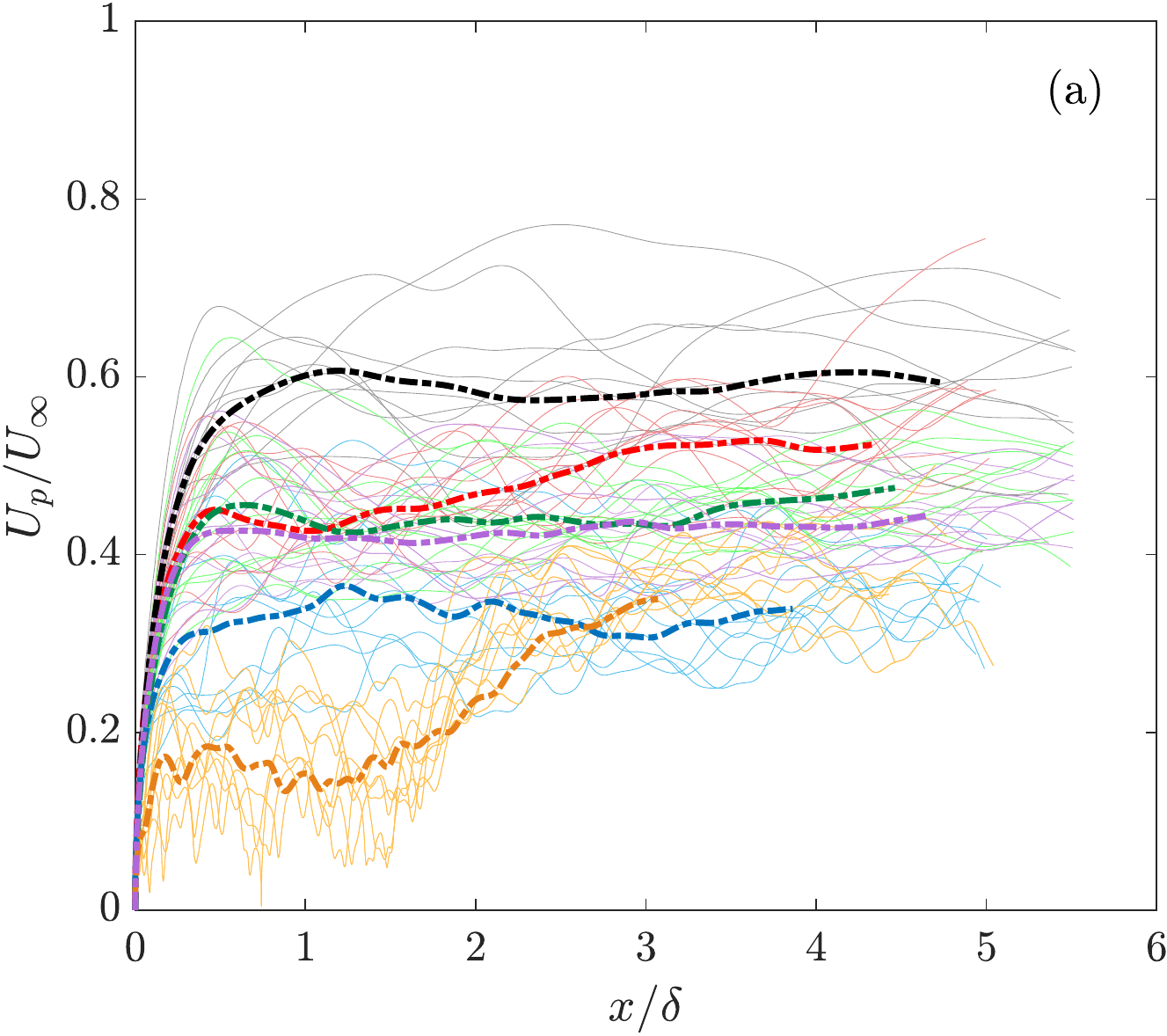}
	}
	\end{subfigure}\hspace{0.5mm}
	\begin{subfigure}{\label{fig:4b}
		\includegraphics[trim={0 1.6mm 0 0},clip,width=.48\textwidth]{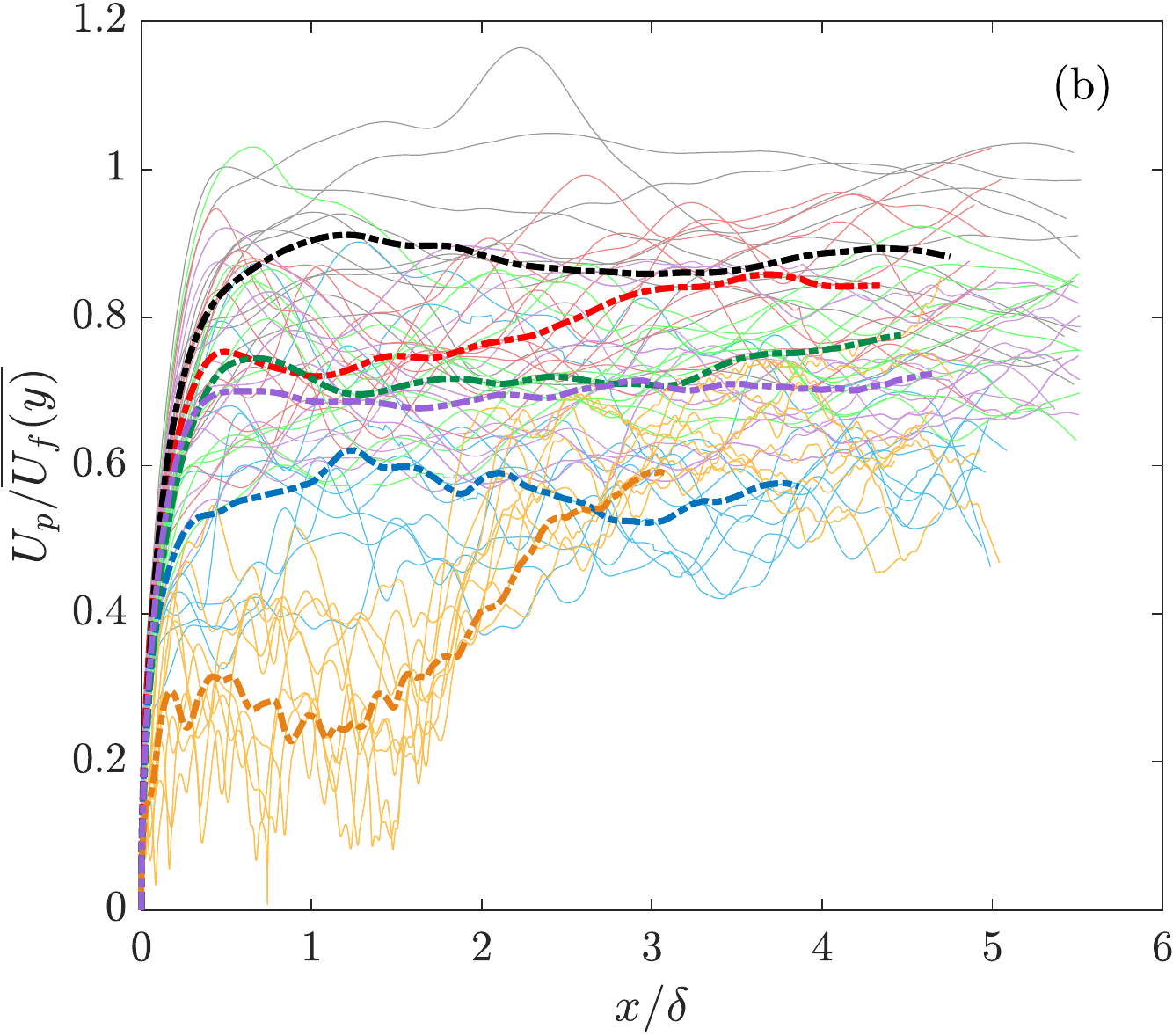}
	}
	\end{subfigure}
		 		\vspace {-3mm}
	\caption{Sphere streamwise velocities, $U_p$ normalized by (a) free-stream velocity, $U_\infty$ and (b) mean unperturbed streamwise fluid velocities at the height of the sphere centroids ($\overline{U(y)}$). Color as in figure \ref{fig:3}.}
	\label{fig:4}
	\vspace{-0.2cm}
\end{figure}

Figure \ref{fig:4} illustrates the sphere streamwise velocities ($U_p$) normalized by the free stream velocity ($U_\infty$) and the mean unperturbed fluid velocities at the height of the sphere centroids ($\overline{U(y)}$). 
Here, $\overline{U(y)}$ is extracted from the mean velocity profile measured by PIV. 
Due to the strong initial mean shear, all spheres accelerated strongly upon release, with the exception of P3 at $Re_{\scriptsize{\slant\tau}}=700$. 
These velocity curves collapsed very well with one another during the initial acceleration phase. 
By contrast, the initial velocity curves of the densest sphere P3 at $Re_{\scriptsize{\slant\tau}}=700$ fluctuated significantly before increasing strongly at $x>1.5\slant\delta$.
As this sphere did not lift off once released, it propagated along the wall with unsteady acceleration.   
It interacted with the wall continuously due to friction.
Friction force is defined as $F_f=fN$, where $f$ is the friction coefficient and $N$ is the normal force.
By approximating $N=F_b-\overline{F_L}$ and assuming a constant friction coefficient in all cases, at $Re_{\scriptsize{\slant\tau}}=700$, $F_f$ for sphere P3 is 16 times larger than for P2.
Hence, when coupled with larger inertia and smaller momentum from the local fluid, the initial acceleration was greatly suppressed.	

In a uniform, steady, unbounded flow, a sphere always accelerates up to zero slip velocity, where terminal velocity is equal to local fluid velocity. 
However, in our studies, the presence of turbulence and the wall have modified the surrounding flow fields and sphere kinematics.
Based on figure \ref{fig:4b}, in most runs, the spheres lagged behind the local fluid.
Sphere P1 reached a higher approximate terminal velocity closer to the mean local flow field than other particles at both $Re_{\scriptsize{\slant\tau}}$. 
For spheres P2 and P3 at $Re_{\scriptsize{\slant\tau}}=700$, the curves began to level off around $0.6 \overline{U_f(y)}$. 
When $Re_{\scriptsize{\slant\tau}}$ was increased, both spheres propagated with higher velocity ($\approx 0.7 \overline{U_f(y)}$), but still significantly slower than the mean flow fields.
Moreover, on some occasions, e.g. sphere P1 at both $Re_{\scriptsize{\slant\tau}}$ and sphere P2 at $Re_{\scriptsize{\slant\tau}}=1300$ traveled faster than the mean local fluid.

As each sphere approached an approximate terminal velocity, velocity fluctuations of order $\pm0.2U_\infty$ were observed in most runs.
The velocity curves also spread over a wide range from the mean curves.
\textcolor{black}{\cite{zeng2008interactions} reported that for a fixed sphere located above the wall, the fluctuation of the time-resolved streamwise force was mainly due to the wall turbulence with only weak influence from vortex shedding.
In their studies, the vortex shedding exhibited stronger effects in the wall-normal and spanwise force components.
In this context, within the region investigated, as the sphere moved away from the wall, the $u_{rms}$ decreased from $0.15$ to $0.10 \overline{U_f(y)}$ at $Re_{\scriptsize{\slant\tau}}=700$ and $0.2$ to $0.1\overline{U_f(y)}$ at $Re_{\scriptsize{\slant\tau}}=1300$, respectively.
This implies that the fast or slow moving zones can accelerate or decelerate the spheres compared with averaged speeds.
As a reference, $U_p/(\overline{U_f(y)}\pm u_{rms})$ of the ensemble-averaged curves could vary $10$-$20\%$ from $U_p/\overline{U_f(y)}$.}

\textcolor{black}{For spheres P2 and P3 at $Re_{\scriptsize{\slant\tau}}=700$, as they interacted with the wall more often, their forward motions were strongly retarded by the friction force as compared to sphere P1 which mostly translated above the wall.
Hence, even after considering the effect of fluctuating velocities, spheres P2 and P3 were still lagging behind the local fluid in most runs.}

\begin{figure}[]

        \centering
        \begin{subfigure}{\label{fig:5a}
        		\includegraphics[trim={0 1.75mm 0 0.21mm},clip,width=.48\textwidth]{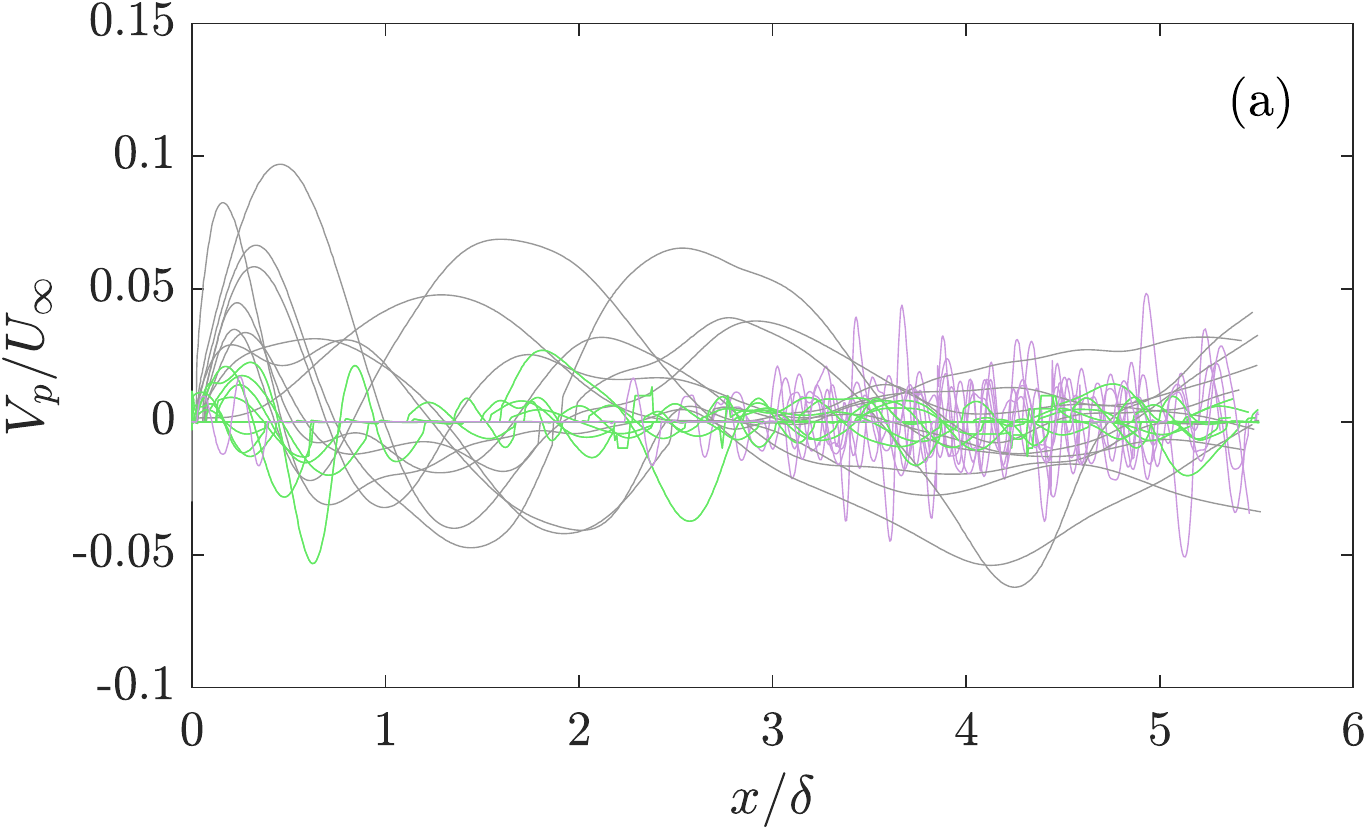}
	}
	\end{subfigure}\hspace{0.2mm}
	\begin{subfigure}{\label{fig:5b}
		\includegraphics[trim={0 1.75mm 0 0.21mm},clip,width=.48\textwidth]{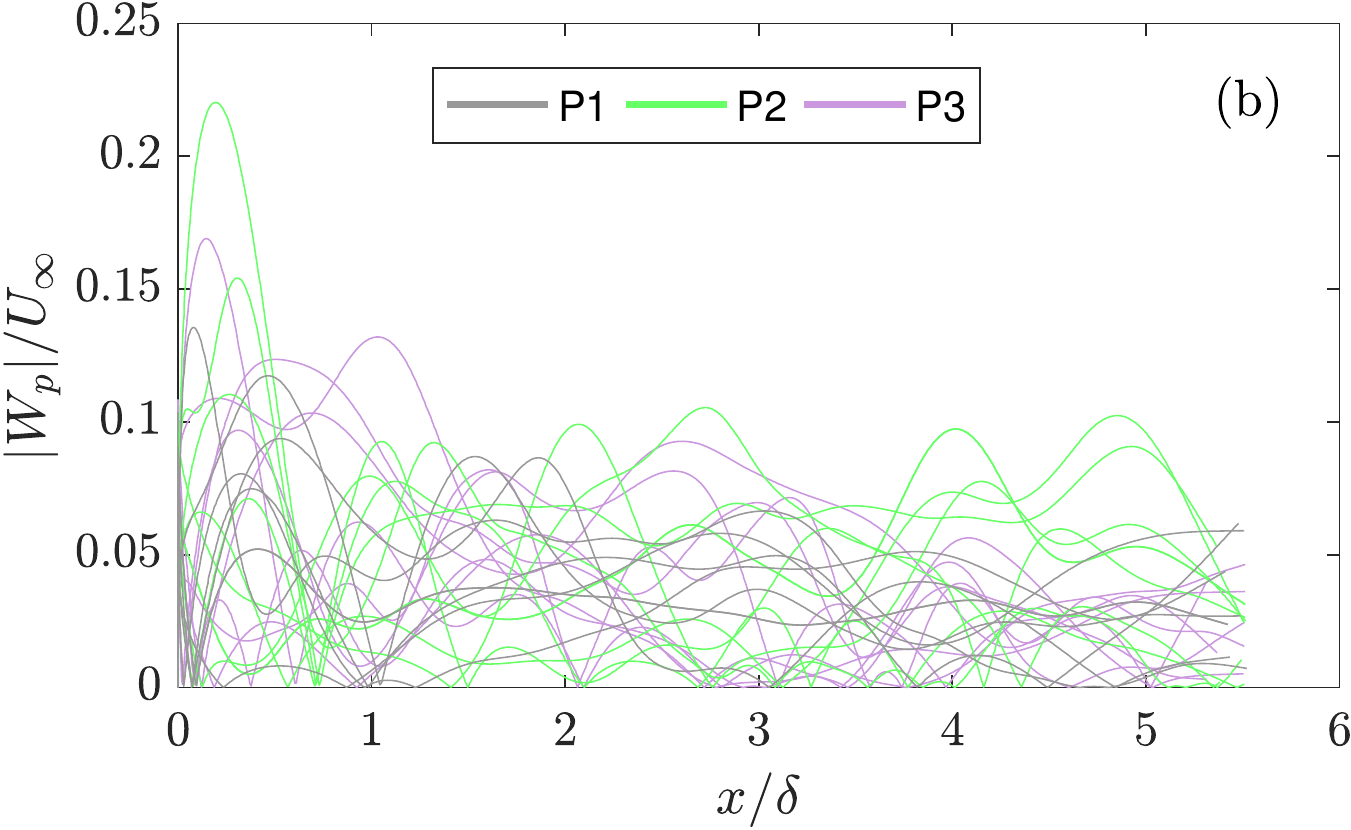}
	}
	\end{subfigure}
	\vspace {-3mm}
	\caption{Sphere (a) wall-normal velocities, $V_p$ and (b) absolute spanwise velocities, $|W_p|$ at $Re_{\scriptsize{\slant\tau}}=1300$, both normalized by free-stream velocity, $U_\infty$.}
		\vspace{-0.2cm}

\end{figure}

The wall-normal velocities ($V_p$) of spheres at $Re_{\scriptsize{\slant\tau}}=1300$ are plotted in figure \ref{fig:5a}. 
The results show that sphere P1 propagated with the highest initial $V_p$, and hence lifted off to a greater height than other lifting spheres.  
At the same time, during most of the lift-off events, sphere P1 traveled with $V_p$ higher than the settling velocity ($|V_s|$) of $0.03U_\infty$ (see Table \ref{tab:1}). 
This signifies a strong upward force opposing the net buoyancy force.
In some instances, the sphere also descended with $V_p$ exceeding $V_s$.
This suggests that aside from the negative buoyancy force, there is an external downward force from the fluid pushing the sphere towards the wall. 
As sphere P2 was denser than P1, it always propagated upward with $V_p$ less than $0.3|V_s|$.
The velocity curves also fluctuated more frequently, indicating that albeit the lift-off magnitudes were smaller, sphere P2 lifted off more often throughout the trajectories.
For sphere P3, as it propagated along the wall downstream, $V_p$ began to increase from zero and fluctuated at an even higher frequency with $V_p<0.16|V_s|$.
These fluctuations corresponded to the  small repeated lift-off events highlighted earlier.
Although P3 lifted off less than $0.25d$ from the wall, the spheres traveled with $V_p$ comparable to those of spheres P1 and P2.
The wall-normal velocity, though smaller than the sphere streamwise velocity, could be accelerated or decelerated by the ejection-sweep events where the sphere gains momentum from the surrounding fluid to propagate in the wall-normal direction.
For instance where the sphere collided with the wall, the coefficient of restitution ($e$), which is the ratio of wall-normal velocity after impact to wall-normal velocity before impact, was computed. 
Among all collision incidents, $e=0$. 
This was because right after colliding with the wall, the sphere slid for a distance of minimum $0.1d$ before lifting off again.
Hence, all collisions were inelastic. 

The absolute spanwise velocities ($|W_p|$) for spheres at $Re_{\scriptsize{\slant\tau}}=1300$ are illustrated in figure \ref{fig:5b}.
Overall, $|W_p|$ varied over larger magnitudes than $|V_p|$.
Most of the time, including during the lift-off events, $|W_p|\geqslant |V_p|$.
Despite the distinct differences in $V_p$ curves and specific gravities, spheres P2 and P3 traveled occasionally with $W_p$ equal to or larger than These results imply that within the buffer and logarithmic layers, the spanwise force is as significant as the wall-normal force \citep[as in][]{zeng2008interactions}.

\begin{figure}[]
        \centering
    \begin{subfigure}{\label{fig:6a}
        		\includegraphics[trim={1 4.5mm 0 1},clip,width=0.29\textwidth]{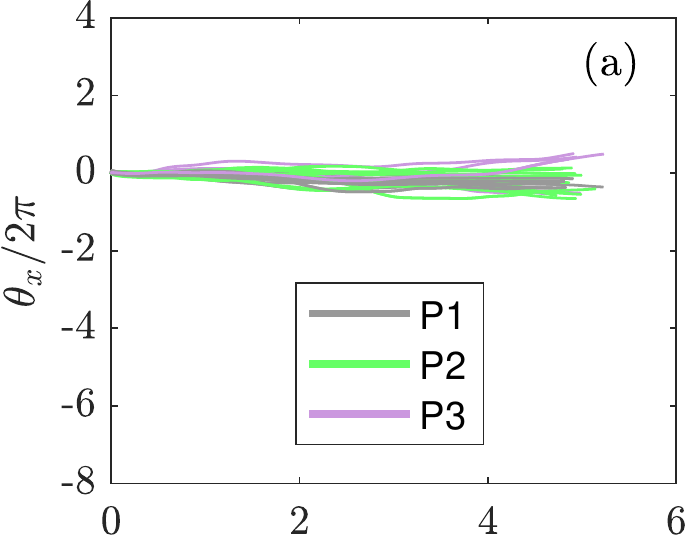}
        		\hspace{0mm}
	}
	\end{subfigure}
	\begin{subfigure}{\label{fig:6b}
	        		\includegraphics[trim={1 4.5mm 0 1},clip,width=0.29\textwidth]{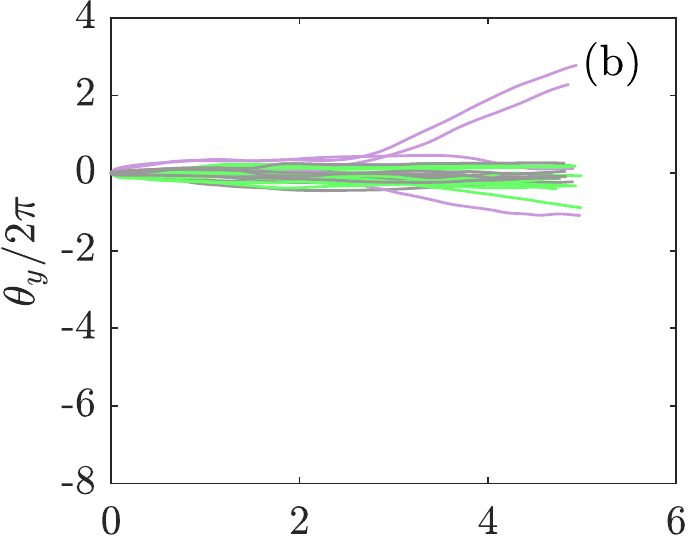}
	        		\hspace{-0.3mm}
	}
	\end{subfigure}
	\begin{subfigure}{\label{fig:6c}
	        		\includegraphics[trim={1 4.5mm 0 1},clip,width=0.29\textwidth]{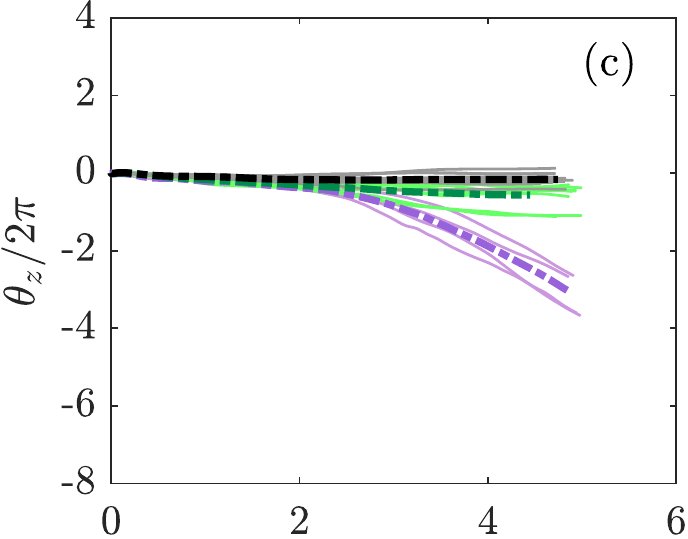}
	        		\hspace{1.2mm}
	}
		\end{subfigure}
	  \begin{subfigure}{\label{fig:6d}
        		\includegraphics[trim={0 0 0.5 2},clip,width=0.29\textwidth]{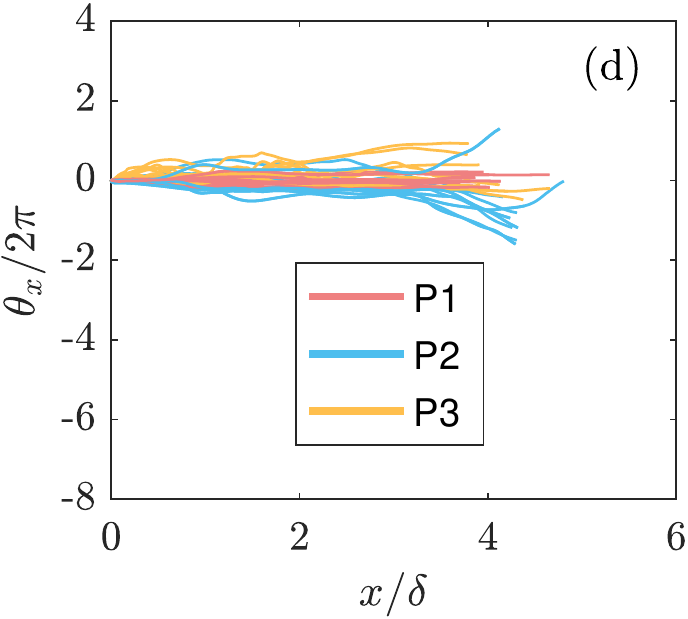}
        		\hspace{0mm}
	}
	\end{subfigure}
	\begin{subfigure}{\label{fig:6e}
	        		\includegraphics[trim={0 0 0.5 2},clip,width=0.29\textwidth]{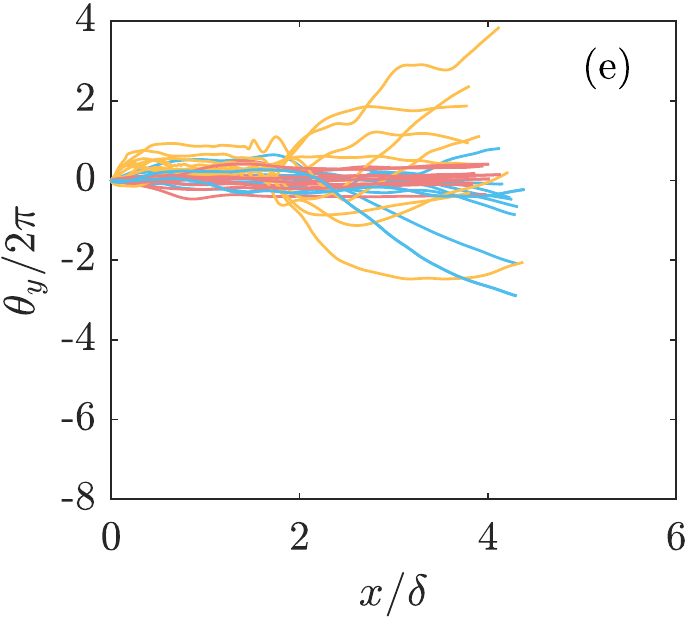}
	        		\hspace{-0.3mm}
	}
	\end{subfigure}
	\begin{subfigure}{\label{fig:6f}
	        		\includegraphics[trim={0 0 0.5 2},clip,width=0.29\textwidth]{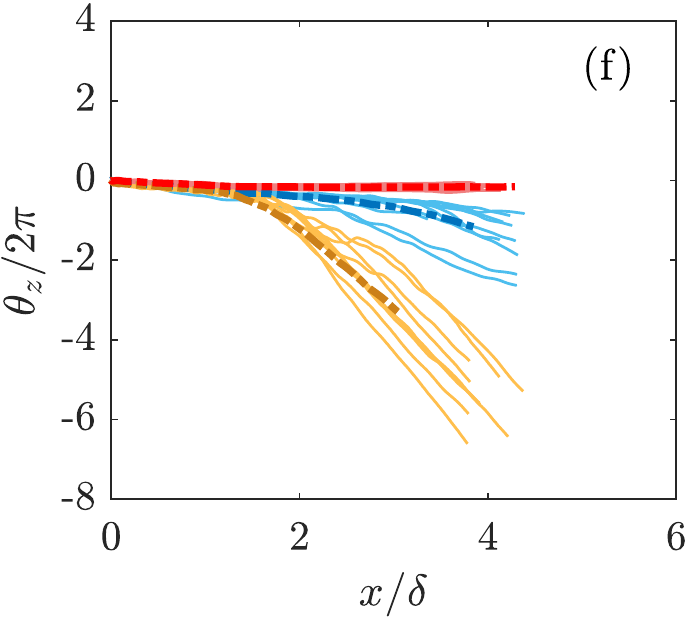}
	}
	\end{subfigure}
	\vspace {-2.5mm}
	\caption{Sphere rotational angle, $\slant{\theta}$ at (a) to (c) $Re_{\scriptsize{\slant\tau}} = 1300$ and (d) to (f) $Re_{\scriptsize{\slant\tau}} = 700$, plotted about streamwise ${x}$-axis, wall-normal ${y}$-axis and spanwise ${z}$-axis, respectively.} 
	\label{fig:6}
		\vspace{-0.05cm}

\end{figure}

\begin{figure}[]
        \centering
        \begin{subfigure}{\label{fig:7a}
        		\includegraphics[trim={0 1.65mm 0 0},clip,width=.48\textwidth]{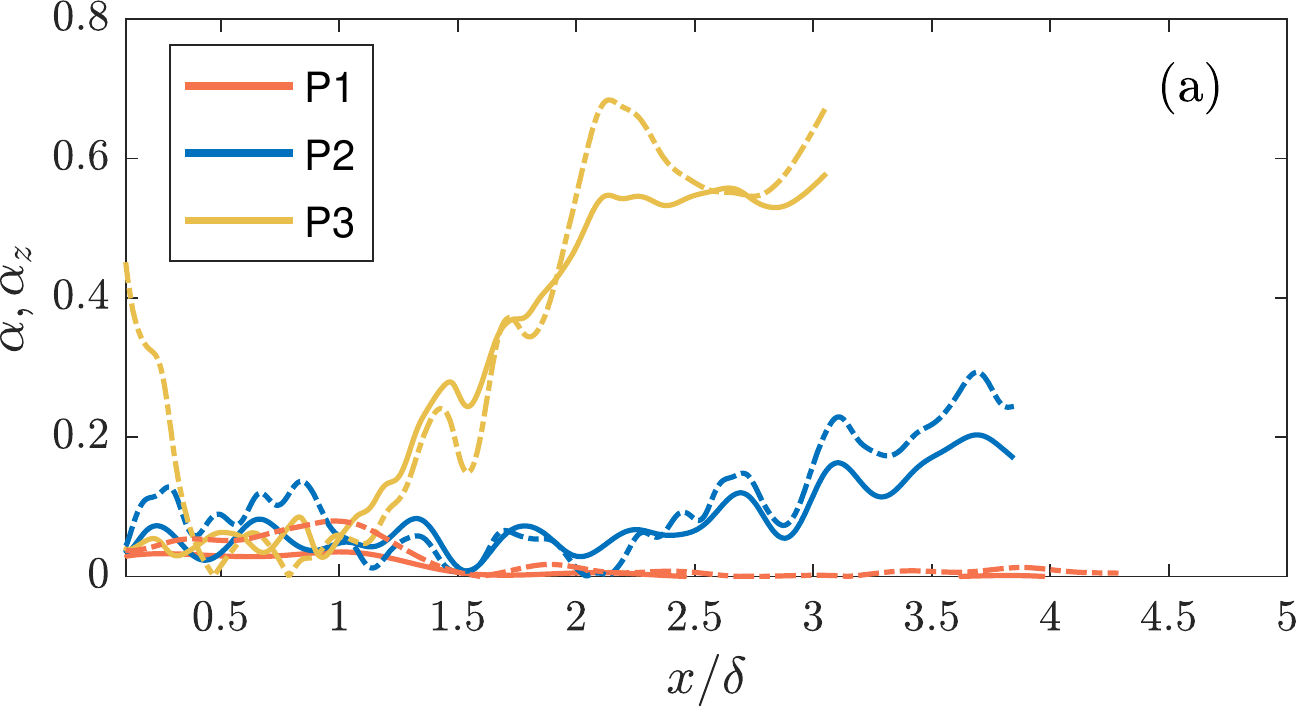}
	}
	\end{subfigure}\hspace{0.2mm}
	\begin{subfigure}{\label{fig:7b}
		\includegraphics[trim={0 1.65mm 0 0},clip,width=.48\textwidth]{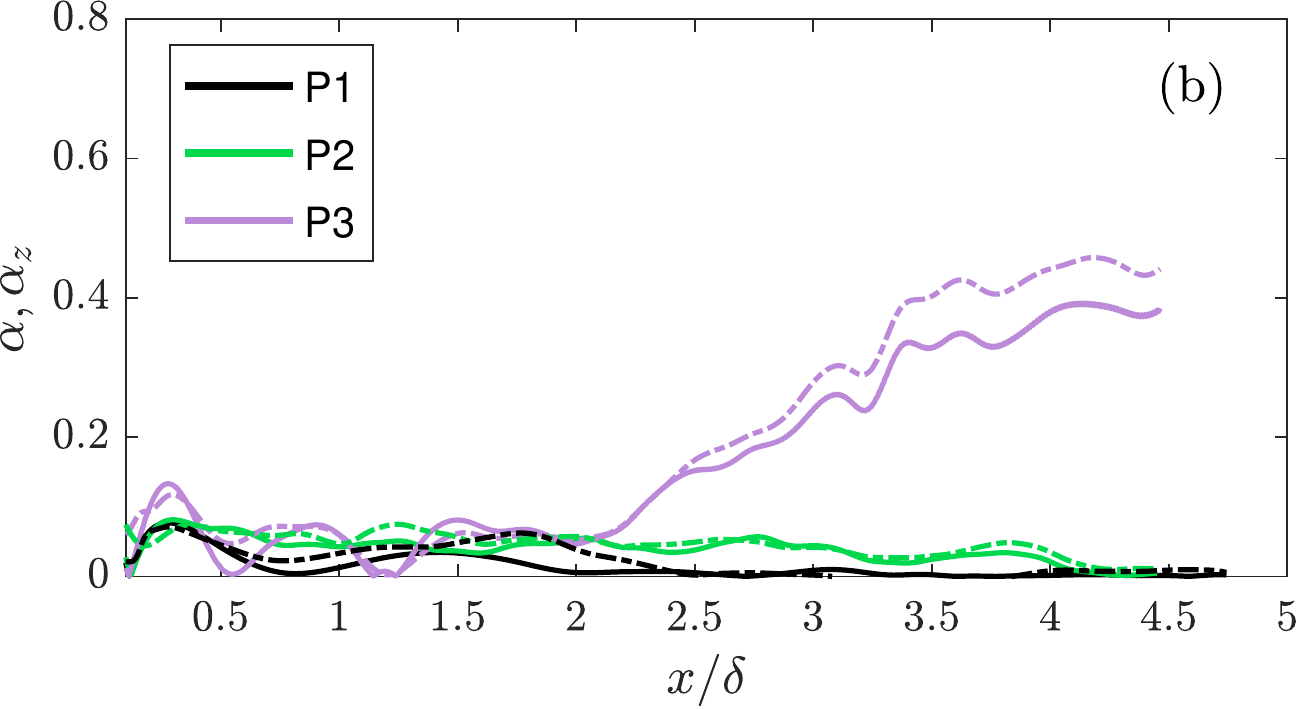}
	}
	\end{subfigure}
	\vspace {-2.9mm}
	 \caption{Ensemble-averaged sphere dimensionless rotation rate at (a) $Re_{\scriptsize{\slant\tau}}=700$ and (b) $Re_{\scriptsize{\slant\tau}}=1300$. Solid lines - $\alpha_z = |\Omega_z|d/2U_p$; Dashed lines - $\alpha = |\bm{\Omega}|d/2|\mathbf{U_p}|$.}
        \label{fig:7}
        	\vspace{-0.15cm}

\end{figure}

We now turn our attention to sphere rotational angle $\slant\theta$  (see figure \ref{fig:6}). 
We also consider two sphere rotation rates ($\Omega$), namely $\alpha_z = |\Omega_z|d/2U_p$ which focuses on rotation about the spanwise axis (forward rotation) to streamwise translation ratio, and $\alpha = |\bm{\Omega}|d/2|\mathbf{U_p}|$ which defines the magnitude of the ratio of rotational velocity to translational velocity. 
In the streamwise direction, if $\alpha_z$ is 0, the sphere is undergoing pure sliding or translating motion without forward spinning; if $\alpha_z$ is 1, the sphere is either undergoing pure forward rolling along the wall without slipping or else forward spinning at the same rate as its streamwise translation. 
The same definition applies to $\alpha$, considering all components of rotation and translation.
The dimensionless rotation rates based on \textcolor{black}{ensemble-averaged} sphere motions are illustrated in figure \ref{fig:7}.

For sphere rotation about the streamwise $x$-axis ($\slant\theta_x$) at $Re_{\scriptsize{\slant\tau}}=1300$, as plotted in figure \ref{fig:6a}, all spheres rotated less than half a revolution. 
The same observation applied to sphere P1 at $Re_{\scriptsize{\slant\tau}}=700$ (see figure \ref{fig:6b}).
After some upstream fluctuations, $\slant\theta_x$ curves stayed very flat. 
When we compared the sphere rotational angle about streamwise axis with their spanwise migration curves, no significant correlations were found in these cases.
By contrast, for spheres P2 and P3 at $Re_{\scriptsize{\slant\tau}}=700$, the $\slant\theta_x$ curves fluctuated with larger magnitude.
As they propagated downstream, $\slant\theta_x$ increased gradually with stronger rotation observed in sphere P2 than P3.
Moreover, these rotations correlated well with the spanwise trajectories in most runs.
\textcolor{black}{In other words, when $\Omega_x$ was positive, $W_p$ was positive; when $\Omega_x$ was negative, $W_p$ was negative.}
These rotations including the spanwise migrations could be initiated by quasi-streamwise vortices, hairpin legs or the spanwise fluid fluctutations which provide the spheres with the required side force and torque to rotate about the $x$-axis and/or to move sideways. 

For rotations about the wall-normal axis ($\slant\theta_y$) and the spanwise $z$-axis ($\slant\theta_z$), two distinct trends were observed. For spheres that mostly traveled above the wall over longer streamwise distances, namely sphere P1 at both $Re_{\scriptsize{\slant\tau}}$ and sphere P2 at $Re_{\scriptsize{\slant\tau}}=1300$, minimal rotations were observed. 
These spheres rotated by less than half a revolution about all axes throughout their propagation.
Notably, even in the presence of strong initial mean shear, the spheres did not develop any significant forward rolling motion (rotation about negative $z$-axis).
These observations were also clearly reflected by the mean dimensionless rotation rate about the spanwise axis where in these cases, $\alpha_z < 0.1$ (see figure \ref{fig:7}). 
When considering the magnitude of rotational velocity, $\alpha$ was slightly larger than $\alpha_z$, but still less than 0.1, signifying weak rotation.
In addition, the mean dimensionless rotation rates depicted that any rotations that took place during the initial phase averaged towards zero further downstream.
Thus, these spheres mainly translated with the fluid while staying away from the wall, or slid along the wall when in contact before lifting off again. 
This also suggests that under the absence of strong rotations, the sphere lift-off events are not due to a Magnus lift force.

By contrast, for spheres that mostly interacted with the wall, namely sphere P2 at $Re_{\scriptsize{\slant\tau}}=700$ and sphere P3 at both $Re_{\scriptsize{\slant\tau}}$, the $\slant\theta_y$ and $\slant\theta_z$ curves changed steeply with increasing $x$.
Upon release, sphere P3 at $Re_{\scriptsize{\slant\tau}}=700$ exhibited a tendency to rotate about the positive $y$-axis.
Then, as this sphere traveled downstream, it rotated continuously about either the positive or negative $y$-axis as indicated by the steep change in $\slant\theta_y$ values.
This rotation could be triggered by adjacent fast and slow moving zones in the $x$-direction which would generate a hydrodynamic torque about $y$-axis.
On the other hand, upon release, $\slant\theta_z$ curves of these non-lifting spheres remained similarly flat as the lifting spheres.
When compared to the rotation about $x$- and $y$-axes, initial $\slant\theta_z$ magnitude was among the least.
In these cases, as no initial lift-off was observed under most instances, the spheres slid along the wall with minimal forward roll. 
After traveling a certain streamwise distance from the origin, the spheres then began to roll forward as indicated by the negative $\slant\theta_z$ slopes.
When considering the mean values of $\alpha$ and $\alpha_z$, both curves increased steeply at various streamwise locations.
The densest sphere P3 rotated with higher velocity at $Re_{\scriptsize{\slant\tau}}=700$ than at $Re_{\scriptsize{\slant\tau}}=1300$, with mean $\alpha_z$ reaching nearly 0.6 and 0.4 respectively.
This implies a significant increase in rolling to sliding tendency as compared to the initial motion. 
Though $\alpha_z$ also increased over time for sphere P2, the sphere mainly slid, reaching $\alpha_z \sim0.2$.
As these spheres rolled forward, they also rotated about the $x$- and the $y$-axes. 
Hence, $\alpha$ typically exceeded $\alpha_z$ after forward rolling has developed.
Interestingly, for sphere P3 at $Re_{\scriptsize{\slant\tau}}=700$, the initial mean $\alpha$ curve deviated significantly from $\alpha_z$. 
This was due to the strong $\slant\theta_y$ when the sphere was first released.
The forward rotations could be initiated by the individual heads of hairpin vortices that move past the spheres or the anti-clockwise vortices about $z$-axis, that provide the spheres with the impulse to roll or spin. 

Lastly, we would like to discuss the small repeated lift-off events observed with sphere P3. 
At $Re_{\scriptsize{\slant\tau}}=700$, we noticed that prior to lifting off at $x\approx2\slant\delta$, the spheres had already begun to roll forward at $x\approx1\slant\delta$.
This was clearly indicated by the increasing value of $\alpha_z$ beginning from $x\approx\slant\delta$ (see figure \ref{fig:7a}).
Then, as the spheres rolled forward, the streamwise velocity curves also increased steeply starting at $x\approx1.5\slant\delta$ (see figure \ref{fig:4}).
This suggests that the delayed streamwise accelerations and thus the repeated lift-off events are possibly prompted by the sphere rotations.
\textcolor{black}{\citeapos{kim2014inverse} experiment on spinning spheres reported that as $\alpha_z$ increased from 0.1 to 0.6, $C_L$ increased from approximately 0.05 to 0.3.
This estimation is comparable to the initial $C_L$ computed based on \citeapos{hall_1988} equation where $C_L=8\overline{F_L}/\pi\slant\rho_f(U_{rel}d)^2 = 0.2$.}
This implies that the contribution of Magnus lift force in these lift-off events may be important. 

\vspace {-1mm}
\section{Conclusions}
\vspace {-0.5mm}

Translation and rotation of spheres with $\slant\rho_s/\slant\rho_f$ of 1.003 (P1), 1.050 (P2) and 1.150 (P3) at $Re_{\scriptsize{\slant\tau}}=700$ and 1300 were successfully reconstructed.
Among all cases, two distinct types of dynamics were observed based on the sphere wall-normal trajectories and rotation behaviors. 
For sphere P1 at both $Re_{\scriptsize{\slant\tau}}$ and sphere P2 at $Re_{\scriptsize{\slant\tau}}=1300$, upon release, the spheres mostly lifted off of the wall and then descended towards the wall due to gravity after reaching an initial peak. 
The spheres either lifted off without returning to the wall or else contacted the wall and then slid along the wall before lifting off again.
Throughout their trajectories, multiple lift-off events including resuspension and saltation with magnitude up to $2d$ from the wall were observed. 
However, no significant rotations developed.
On the other hand, for sphere P2 at $Re_{\scriptsize{\slant\tau}}=700$  and sphere P3 at both $Re_{\scriptsize{\slant\tau}}$, in most runs, the spheres did not lift off once released.
Instead of rolling forward, they slid along the wall with some minor rotations about the $x$- and $y$-axes. 
As the spheres propagated downstream, forward rolling as well as occasional small lift-off events of magnitude less than $0.25d$ began to develop. 
Here, Magnus lift may be important for these forward spinning spheres.
In terms of spanwise migration, all spheres traveled with spanwise velocities equal to or larger than the wall-normal velocities. 
This implies that the spanwise force is as significant as the wall-normal force.

In the context of particle-wall interactions, the presence of the wall altered the pressure distribution around a sphere in contact with it.
The sphere experienced a positive lift force and lifted off of the wall when $F_L>F_b$. 
However, the presence of wall friction also retarded the spheres from accelerating towards the local fluid velocity.
This was especially true for sphere P3 at $Re_{\scriptsize{\slant\tau}}=700$ where instead of accelerating strongly as other spheres did upon release, this sphere slid along the wall with unsteady acceleration.
In all cases, the spheres were seen lagging behind the fluid even after attaining an approximate terminal velocity.
Meanwhile, for all instances where the spheres collided with the wall, collisions were inelastic. 

Our results also suggest the presence of significant particle-turbulence interactions within the boundary layer. 
The fluctuations observed in the streamwise velocity curves and also the rotation about wall-normal axis could be initiated by the slow and fast moving zones that the spheres encountered locally while propagating.
The lift-off events and thus the wall-normal velocity could be driven by the ejection-sweep events where the spheres gained momentum to ascend to a greater height or descend with velocity larger than the settling velocity.
Meanwhile, the quasi-streamwise vortices and the hairpin legs can also lead to spanwise migration, spanwise velocity fluctuations as well as the rotation about streamwise axis which were strongly correlated with each other in some cases.
The individual heads of hairpin vortices or the anti-clockwise vortices about the spanwise axis may also be responsible for the development of the forward rotation downstream of the sphere propagation.
{Additionally, as the spheres were propagating into their own wakes, and the particle Reynolds numbers were high, the effect of vortex shedding could be important.}

Within the turbulent boundary layer, the cases studied have shown distinctive behaviors due to both friction and coherent structures. 
To complement our observations, experiments that quantify the velocity fields surrounding a sphere will be conducted next to better understand the particle-wall and particle-turbulence interactions.
\begin{acknowledgments}
\vspace {-0.5mm}
The authors thank Nicholas Morse, Ben Hiltbrand and Alessio Gardi for their help with particle fabrication and image processing. This work is funded by National Science Foundation (NSF CBET-1510154).
\end{acknowledgments}
\vspace {-1mm}
\bibliography{ISPIV2019_References}

\end{document}